# Demonstration of improved sensitivity of echo interferometers to gravitational acceleration

C. Mok,[1] B. Barrett,[1,*] A. Carew,[1] R. Berthiaume,[1] S. Beattie,[1,†] and A. Kumarakrishnan[1,‡]

[1]*Department of Physics and Astronomy, York University,
4700 Keele Street, Toronto, Ontario M3J 1P3, Canada*

(Dated: June 7, 2013)

We have developed two configurations of an echo interferometer that rely on standing wave excitation of a laser-cooled sample of rubidium atoms. Both configurations can be used to measure acceleration $a$ along the axis of excitation. For a two-pulse configuration, the signal from the interferometer is modulated at the recoil frequency and exhibits a sinusoidal frequency chirp as a function of pulse spacing. In comparison, for a three-pulse stimulated echo configuration, the signal is observed without recoil modulation and exhibits a modulation at a single frequency as a function of pulse spacing. The three-pulse configuration is less sensitive to effects of vibrations and magnetic field curvature leading to a longer experimental timescale. For both configurations of the atom interferometer (AI), we show that a measurement of acceleration with a statistical precision of 0.5% can be realized by analyzing the shape of the echo envelope that has a temporal duration of a few microseconds. Using the two-pulse AI, we obtain measurements of acceleration that are statistically precise to 6 parts per million (ppm) on a 25 ms timescale. In comparison, using the three-pulse AI, we obtain measurements of acceleration that are statistically precise to 0.4 ppm on a timescale of 50 ms. A further statistical enhancement is achieved by analyzing the data across the echo envelope so that the statistical error is reduced to 75 parts per billion (ppb). The inhomogeneous field of a magnetized vacuum chamber limited the experimental timescale and resulted in prominent systematic effects. Extended timescales and improved signal-to-noise ratio observed in recent echo experiments using a non-magnetic vacuum chamber suggest that echo techniques are suitable for a high precision measurement of gravitational acceleration $g$. We discuss methods for reducing systematic effects and improving the signal-to-noise ratio. Simulations of both AI configurations with a timescale of 300 ms suggest that an optimized experiment with improved vibration isolation and atoms selected in the $m_F = 0$ state can result in measurements of $g$ statistically precise to 0.3 pbb for the two-pulse AI and 0.6 ppb for the three-pulse AI.

## I. INTRODUCTION

The sensitivity of cold atom interferometers (AIs) to gravitational acceleration, $g$, has stimulated efforts to understand the nature of this fundamental interaction as well as development of sensitive portable instruments for practical applications. Measurements related to basic science include the determinations of the universal gravitational constant [1, 2], tests of Lorentz invariance [3], and proposed tests of general relativity [4]. Practical applications of AIs include geodesic surveys connected with oil, natural gas and mineral exploration, as well as the accurate determination of tidal variations. The development of portable sensors to realize this goal has been the focus of extensive research [5–15].

The most precise measurements of $g$ are derived from superconducting quantum interference based devices [16, 17], whereas the best portable industrial measurements of $g$ are obtained using a falling corner-cube Mach-Zehnder interferometer [5] with a drop height of 0.3 m. This type of optical interferometer achieves an absolute accuracy of 1 part per billion (ppb) in an integration time of 20 minutes.

Recent improvements in cold atom-based sensors began with the pioneering experiments in Refs. [18, 19] which relied on a Raman interferometer to achieve a statistical precision of 3 parts per million (ppm) in a measurement time of 1000 seconds. The Raman AI relies on the manipulation of cold atoms between two hyperfine ground states using optical transitions. This AI uses velocity selected atoms that are optically pumped into an $m_F = 0$ sublevel to reduce magnetic effects. Experiments based on the Raman AI have resulted in the most sensitive atom-based measurements of $g$. In Ref. [3], a statistical precision of 1.3 ppb was achieved in 75 seconds of data acquisition, whereas Ref. [20] included a detailed study of systematic effects and reported a statistical precision of 3 ppb in 1 minute of interrogation time. Both experiments relied on active vibration stabilization of the inertial reference frame. In Refs. [3, 20], the atoms were launched in a 50 cm atomic fountain, resulting in a free-fall measurement time of 320 ms. More recently, the Raman AI has been implemented with a 6.5 m drop zone to achieve an inferred single shot sensitiv-ity of $7 \times 10^{-12}$ [21]. The Raman AI has also produced the best atom-based measurements of gravity gradients [22, 23] and rotations [21, 24, 25]. As a result, this AI has been extensively developed for remote sensing [6–8, 12–15].

---

[*] Present address: Laboratoire Photonique Numérique et Nanosciences, Institut d'Optique d'Aquitaine, Université Bordeaux 1, 33405 Talence, France
[†] Present address: Department of Physics, University of Toronto, Toronto, Ontario M5S 1A7, Canada
[‡] akumar@yorku.ca

Alternative techniques using Bloch oscillations [26, 27] have shown potential for realizing more compact setups. In particular, the work in Ref. [27] has measured $g$ with a statistical precision of 100 ppb. However, in this experiment, the accuracy relies on the precise knowledge of $h$, which is known to 44 ppb [28]. More recently, a hybrid technique involving the Raman AI with large momentum transfer Bragg pulses has reported a sensitivity of 2.7 ppb with 1000 seconds of data acquisition using a drop height of 20 cm and passive vibration stabilization [29].

In this paper, we describe sensitive measurements of the total acceleration, $a$, along the axis of excitation using two configurations of a single-state, time-domain AI [30–34] that rely on echo techniques and utilized samples of laser-cooled Rb atoms released from a magneto-optical trap (MOT). The echo AI, which is sensitive to the absolute acceleration, uses a single excitation frequency and does not require velocity selection. Here, we compare and contrast the characteristics of the two-pulse AI and three-pulse stimulated-echo AI for measurements of $a$ [35].

We first discuss the physical principles of the two AI configurations using recoil diagrams.

## A. Physical Principles

Figure 1(a) represents the recoil diagram for the two-pulse configuration of the AI in the absence of gravity. A sample of laser-cooled Rb atoms is excited along the vertical by two standing wave (sw) pulses separated by a time $T_{21}$. The sw pulses are composed of two traveling wave components, each having a wave vector of magnitude $k$. Atoms in all magnetic sublevels of the $F = 2$ ground state of $^{87}$Rb or $F = 3$ ground state of $^{85}$Rb manifolds are diffracted into a superposition of momentum states separated by $\hbar q$ at $t = T_1$. Here, $q = 2k$ for counter-propagating traveling wave components of the sw excitation so that the wavelength of the optical potential is $\lambda/2$.

For an echo AI, the phases of momentum states corresponding to the same internal ground state are modulated in time as $n^2 \omega_q t$ due to the recoil of atoms that coherently scatter $n$ photons from standing wave excitation pulses. Here, the recoil frequency is $\omega_q = \hbar q^2/(2M)$, where $M$ is the atomic mass. The interference between momentum states produces a one dimensional density grating with a period of $\lambda/2$ immediately after each excitation pulse. The initial velocity distribution of the cold sample along the sw axis causes the grating to dephase on a timescale of $\tau_{\rm coh} = 1/(ku)$ where $u = \sqrt{2k_B \mathcal{T}/M}$ is the $1/e$ width of the velocity distribution. Here, $\mathcal{T}$ is the temperature of the sample, and $k_B$ is the Boltz-mann constant. The coherence time ($\tau_{\rm coh} \sim 2$ $\mu$s for a 20 $\mu$K sample) is characteristic of the time scale required for an atom in the sample to traverse one period of the sw potential. This timescale is considerably shorter than the recoil period $\tau_q = \pi/\omega_q = 33.655$ $\mu$s for $^{87}$Rb. The AI uses an echo technique to cancel the effect of velocity dephasing and observe the recoil modulation by exciting the sample with a second sw pulse at $t = T_{21}$. As a result, momentum states separated by $\hbar q$ interfere near the echo time $t_{\rm echo} = 2T_{21}$, giving rise to a rephased density grating.

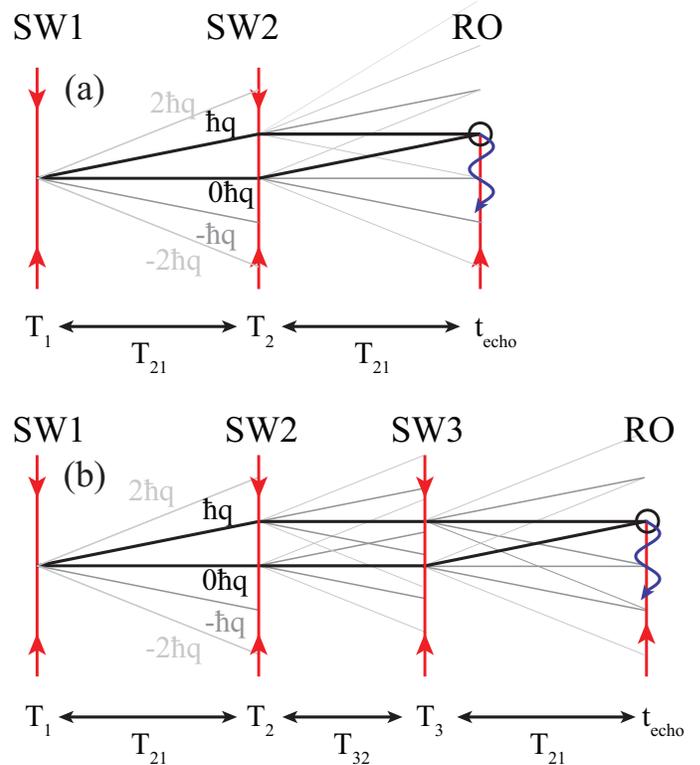

FIG. 1. (Color online) Recoil diagrams for (a) two-pulse and (b) three-pulse AIs in the absence of gravity. Only a subset of all trajectories are shown. SW refers to standing wave pulses and RO is a traveling wave read-out pulse. The sw pulses, composed of two counter-propagating traveling wave components, each with wave vector $|\mathbf{k}|$, diffract atoms into a superposition of momentum states separated by $\hbar q$. For both two-pulse and three-pulse AIs, the backscattered signal arises from interferences between states differing by $\hbar q$ at the echo time.

The grating contrast and phase are measured in the vicinity of the echo time by applying a traveling wave read-out (RO) pulse and detecting the electric field amplitude and phase of the coherently backscattered light. The backscattered electric field during the RO pulse is known as the echo envelope. The temporal duration of the echo envelope is $\tau_{\rm coh}$ while the amplitude of the echo envelope is modulated at the recoil frequency as a function of $T_{21}$. The recoil period is the time in which an atom that has acquired a recoil velocity of $\hbar q/M$ traverses a distance $\lambda/2$, so that $\tau_q = \dfrac{\lambda/2}{\hbar q/M}$. The contrast of the density grating is modulated at this period. For precise measurements of $\omega_q$, the modulation of either the electric field or intensity of the backscattered light is measured at the echo time as a function of $T_{21}$ [30–32, 36–38].

In contrast, a precise determination of $g$ requires a measurement of both the amplitude and phase of the falling grating as a function of the pulse separation $T_{21}$. The essential characteristic of the two-pulse AI is that the different momentum states comprising the arms of the interferometer experience a continuously changing relative displacement during the entire experiment. In the presence of gravitational acceleration, the recoil modulated signal as a function of $T_{21}$ acquires an additional frequency chirped, sinusoidal modulation. The corresponding increase in the phase of the AI scales as $qgT_{21}^2$ due to the free-fall of atoms. Although this aspect is appealing for a precision measurement of $g$, the signal amplitude exhibits recoil modulation as well as a chirped frequency as a function of $T_{21}$, resulting in the need for a complicated fit function to extract $g$. The signal from the AI is analogous to the interference fringes recorded by the falling corner-cube optical interferometer in Ref. [5].

An alternate, "stimulated-echo" configuration that involves three sw pulses is shown in Fig. 1(b). Here, the first sw pulse creates a superposition of momentum states separated by $\hbar q$. A second sw pulse applied at $t = T_{21}$ results in momentum states that are co-propagating at fixed separation with the same momentum. A third pulse applied at $t = T_{21} + T_{32}$ causes the co-propagating states to interfere at the echo time $t_{\text{echo}} = 2T_{21} + T_{32}$, forming a density grating. As in the two-pulse AI, the grating formation is associated with interference of momentum states separated by $\hbar q$.

The stimulated echo configuration was first developed in Ref. [39] for collisional studies. Subsequently, an AI based on this configuration was used to study the formation of nanostructures in cold atoms [40]. Reference [33] used an inclined magnetic waveguide to show that this AI exhibits reduced sensitivity to mirror vibrations for measurements of acceleration. The theory of this interferometer was described in detail in Ref. [34] and predictions were verified using measurements of magnetic field gradients that did not require vibration stabilization.

The arms of the three-pulse AI consist of co-propagating wavepackets with no momentum difference during the central time window associated with $T_{32}$ [33, 34]. The amplitude of the echo envelope as a function of pulse separation $T_{32}$ shows no recoil modulation. However, the in-phase and in-quadrature component amplitudes of the echo are modulated at a fixed frequency determined by $T_{21}$. This angular frequency, which can be shown to be $qgT_{21}$, is determined by the velocity $gT_{21}$ the atoms acquire during the time interval $T_{21}$ due to gravity. Therefore, this configuration represents a velocity sensitive AI. The period of the in-phase and in-quadrature component amplitudes as a function of $T_{32}$ is given by

$\tau_{\text{v}} = \dfrac{\lambda/2}{gT_{21}}$.

The absence of recoil modulation and the constant modulation period $\tau_{\text{v}}$ have beneficial practical consequences, improving the quality of the fits to the data, which results in increased precision. Since there is no relative displacement between the co-propagating momentum states during $T_{32}$, the AI exhibits reduced sensitivity to spurious accelerations due to magnetic gradients and mirror vibrations [33]. This feature allows the total timescale of the experiment to be increased in comparison to the two-pulse AI [34], further improving the precision. By careful choice of pulse parameters, a suitably long observational window during which $T_{32}$ is varied is made available for the measurement of the fixed frequency. In this manner, the time interval $T_{21}$ during which the AI is sensitive to spurious accelerations can be minimized. The disadvantages of this technique include the reduction of the signal amplitude due to the additional standing wave interaction and the sensitivity of gravity measurements to any initial velocity along the sw axis.

In this paper, we use both AI configurations to show that the echo envelope, with a duration of a few microseconds, can be used to extract $a$ with a statistical precision of 0.5%. The amplitude and phase of the two-pulse echo is analyzed as a function of $T_{21}$ to demonstrate a measurement of $a$ with a statistical precision of 6 ppm. The timescale of this measurement ($\approx 25$ ms) is limited by the residual magnetization of the stainless steel vacuum chamber. We show that the three-pulse AI can be used to increase the total timescale to $\approx 50$ ms and realize a statistical precision of 75 ppb. These results were obtained using an apparatus in which only critical optical components were passively isolated from vibrations. Additionally, since no magnetic state selection was performed, all magnetic sub-levels of the ground state manifold contributed to the signal.

The rest of the paper is organized as follows. In section II, we review expressions for the atom interferometric phase due to gravity for both AIs, and discuss relevant features of the echo envelope and amplitude. Section III provides an overview of the experimental setup. In section IV, we discuss different methods of extracting $a$. The paper concludes with section V, which contains a discussion of systematic effects pertaining to a precise measurement of gravitational acceleration, and prospects for future work.

## II. THEORY

We now review the theoretical description of both the two- and three-pulse AI configurations, and the characteristics of their corresponding signals. Both two-pulse and three-pulse AIs involve excitation of a laser-cooled sample by sw pulses applied along the vertical. The traveling-wave components of the sw pulses are detuned with respect to the excited state so that the effects of spontaneous emission are reduced [36]. The sw pulse durations $\tau_1, \tau_2, \tau_3$, are sufficiently short such that the motion of the atoms during the excitation can be ignored (Raman-Nath regime). Under these conditions, the phase accumulation due to gravity in the two-pulse [30–32] and three-pulse AIs [34] has been extensively discussed on the basis of a quantum mechanical approach. Whereas Ref. [32] considered a single magnetic sublevel, the extension of the theoretical treatment to a manifold of ground state magnetic sublevels is discussed in Refs. [31, 34, 41]. Here, we present a set of simplified equations that apply to an atomic system with a single magnetic ground state sublevel. This simplified treatment is sufficient for understanding the signal characteristics. For the theoretical treatment that follows, $g$ represents a constant gravitational acceleration along the axis of sw excitation.

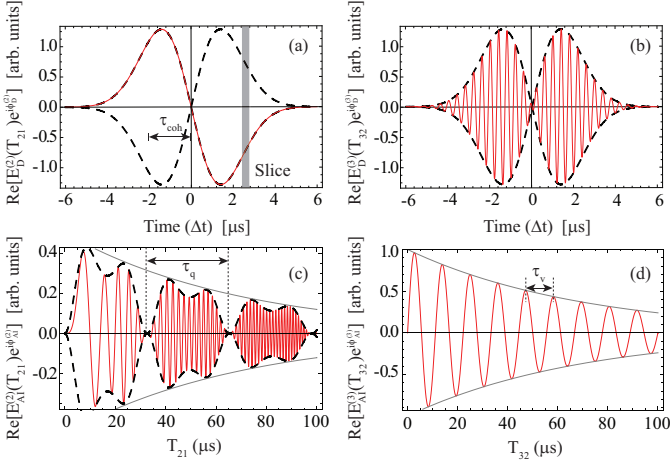

FIG. 2. (Color online) Predicted shape of the in-phase component of the echo envelope at (a) $2T_{21} = 0.1$ ms for the two-pulse AI and (b) $2T_{21} + T_{32} = 100$ ms for the three-pulse AI are shown as a solid red lines. Here $g = 9.8$ m/s$^2$, and $\Delta t$ is the time measured with respect to the echo time. In (a), a time slice used in the analysis is shown as a gray rectangle. The two echo envelopes shown in red are described by $E_D^{(2)} e^{i\phi_D^{(2)}}$ for the two-pulse AI (see Eqs. 5 and 7), and $E_D^{(3)} e^{i\phi_D^{(3)}}$ for the three-pulse AI (see Eqs. 14 and 16). The black dashed lines show the envelopes in the absence of gravity. The echo envelope exhibits an increase in the oscillation frequency as the free-fall time increases. (c) Predicted shape of the in-phase component of the signal amplitude for the two-pulse AI as a function of $T_{21}$ as predicted by $E_{AI}^{(2)} e^{i\phi_{AI}^{(2)}}$ (see Eqs. 6 and 8) and shown as a solid red line. The signal exhibits chirped sinusoidal behaviour due to the quadratic dependence of $\phi_{AI}^{(2)}$ on $T_{21}$ and also shows recoil modulation with a period $\tau_q = 33.151$ $\mu$s. Here we set $g = 980$ m/s$^2$ for illustrative purposes. The black dashed lines show the total signal amplitude. (d) Predicted shape of the in-phase component of the signal amplitude for the three-pulse AI as a function of $T_{32}$ as predicted by $E_{AI}^{(3)} e^{i\phi_{AI}^{(3)}}$ (see Eqs. 15 and 17) and shown as a solid red line. The signal exhibits a constant modulation frequency with a period $\tau_v$ and shows no recoil modulation. Here $T_{21} = 15$ ms. Again, we set $g = 980$ m/s$^2$ for illustrative purposes. The total signal amplitudes for both two-pulse and three-pulse AIs show a phenomenological exponential decay (gray line) to illustrate signal loss due to decoherence and transit time effects.

## A. Two-Pulse AI

The backscattered electric field for the two-pulse AI can be written as

$$E_g^{(2)} = E_0^{(2)} e^{i\phi_g^{(2)}}, \qquad (1)$$

where $E_0^{(2)}$ is the electric field amplitude and $\phi_g^{(2)}$ is the gravitational phase. The electric field amplitude for the two-pulse AI can be shown to be

$$\begin{aligned} E_0^{(2)} \propto\ &E_{RO} e^{-(\Delta t/\tau_{\rm coh})^2} J_1 \left[2\theta_1 \sin(\omega_q \Delta t)\right] \\ &\times J_2\{2\theta_2 \sin\left[\omega_q(T_{21} + \Delta t)\right]\} \\ &\times e^{-t_{\rm echo}/\tau_{\rm decay}}, \end{aligned} \qquad (2)$$

where $E_{RO}$ is the electric field amplitude of the read-out pulse, $J_n(x)$ is the $n$th-order Bessel function of the first kind, $\theta_1$ and $\theta_2$ are the pulse areas of the first and second sw pulses respectively, $\Delta t = t - 2T_{21}$ is the time relative to the echo time $t_{\rm echo} = 2T_{21}$, and $\omega_q = \hbar q^2/(2M)$ is the two-photon recoil frequency. Here, $\tau_{\rm coh} = 1/ku$ is the coherence time due to Doppler dephasing that defines the temporal width of the signal shown in Fig. 2(a), where $u = \sqrt{2k_B \mathcal{T}/M}$ is $1/e$ width of the one dimensional velocity distribution along the excitation beams and $\mathcal{T}$ is the temperature of the laser-cooled sample. The last term in Eq. 2 represents a phenomenological decay, with a time constant $\tau_{\rm decay}$ that models the effect of signal loss due to all decoherence mechanisms as well as the transit time of cold atoms through the interaction zone defined by the excitation beams.

As shown in Refs. [31, 32, 34], in the presence of gravity, the phase of the two-pulse AI is given by

$$\phi_g^{(2)} = -\frac{qg}{2}(2T_{21}^2 + 4T_{21}\Delta t + \Delta t^2). \qquad (3)$$

This phase is proportional to the space-time area enclosed by the interferometer pathways.

It is useful to decouple the expression for the signal into two parts that are dependent on the timescales $T_{21}$ and $\Delta t$ to explain the characteristics of the echo signal. Therefore we write the two-pulse echo signal as

$$E_g^{(2)} = E_D^{(2)}(\Delta t) E_{AI}^{(2)}(T_{21}) e^{i\phi_D^{(2)}(\Delta t)} e^{i\phi_{AI}^{(2)}(T_{21})}, \qquad (4)$$

where

$$\phi_D^{(2)}(\Delta t) = -qg(2T_{21}\Delta t + \Delta t^2/2), \qquad (5)$$

is the Doppler phase which is dependent on $\Delta t$, and

$$\phi_{AI}^{(2)}(T_{21}) = -qgT_{21}^2 \qquad (6)$$

is the AI phase which is dependent on only $T_{21}$. The Doppler electric field amplitude is given by

$$E_{\rm D}^{(2)}(\Delta t) \propto E_{\rm RO} e^{-(\Delta t/\tau_{\rm coh})^2} J_1\left[2\theta_1 \sin(\omega_q \Delta t)\right], \quad (7)$$

and in the limit $\Delta t \ll T_{21}$, the AI electric field amplitude is given by

$$E_{\rm AI}^{(2)}(T_{21}) = J_2\{2\theta_2 \sin\left[\omega_q T_{21}\right]\} e^{-t_{\rm echo}/\tau_{\rm decay}}. \quad (8)$$

The measurement of gravity is based on detecting the amplitude and phase of the light back-scattered by the atomic grating (which has a frequency $\omega_{\rm AI}$ and phase $\phi_{\rm AI}$) with reference to an optical local oscillator (LO), which has a fixed frequency $\omega_{\rm LO}$ and phase $\phi_{\rm LO}$. The detection system records the signal in the form of a beat note at the frequency $\omega_{\rm AI} - \omega_{\rm LO}$ and with a phase difference $\phi_{\rm signal} = \phi_{\rm AI} - \phi_{\rm LO}$. The phase shifts associated with the atoms measured using this detection technique is sensitive to optical phase shifts of the sw pulses and the LO due to the environment. This method of detection allows the in-phase and in-quadrature components $E_0^{(2)} \cos(\phi_g^{(2)})$ and $E_0^{(2)} \sin(\phi_g^{(2)})$ to be recorded. The total signal amplitude is

$$E_0^{(2)} = \frac{1}{\sqrt{2}} \left\{ \left[E_0^{(2)} \cos(\phi_g^{(2)})\right]^2 + \left[E_0^{(2)} \sin(\phi_g^{(2)})\right]^2 \right\}^{1/2}. \quad (9)$$

It is possible to remove the recoil-modulation and signal-decay terms from the in-phase and in-quadrature components of the back-scattered field amplitude by normalizing with respect to $E_0^{(2)}$. Then, we are left with $\cos(\phi_g^{(2)})$ and $\sin(\phi_g^{(2)})$ as the two components of the signal.

The dashed lines in Figs. 2(a) show the Doppler electric field amplitude as predicted by Eq. 7. This shape is generated by choosing a convenient $T_{21}$ to maximize the recoil modulated signal, modeled by Eq. 8. The solid red line shows oscillations within the echo envelope due to gravity as predicted by $E_{\rm D}^{(2)} e^{i\phi_{\rm D}^{(2)}}$. The oscillations are due to the atoms falling through a grating spacing of $\lambda/2$, resulting in a phase increment of $2\pi$. This effect can also be described as a Doppler shift of the backscattered field due to the motion of the falling grating.

The solid red line in Fig. 2(c) shows the predicted shape of the in-phase component of the signal amplitude for the two-pulse AI as a function of $T_{21}$ as predicted by $E_{\rm AI}^{(2)} e^{i\phi_{\rm AI}^{(2)}}$ (see Eqs. 6 and 8). This signal shows a recoil modulation and a frequency chirped oscillation due to gravity. The black dashed line shows the recoil modulated total signal amplitude $E_{\rm AI}^{(2)}$.

### B. Three-Pulse AI

The backscattered electric field for the three-pulse AI can be written as

$$E_g^{(3)} = E_0^{(3)} e^{i\phi_g^{(3)}}, \quad (10)$$

where $E_0^{(3)}$ is the electric field amplitude and $\phi_g^{(3)}$ is the gravitational phase. The electric field amplitude for the three-pulse AI can be shown to be [34]:

$$\begin{aligned}E_0^{(3)} \propto & E_{\rm RO} e^{-(\Delta t/\tau_{\rm coh})^2} J_1\left[2\theta_1 \sin(\omega_q \Delta t)\right] \\ & \times J_1\{2\theta_2 \sin\left[\omega_q (T_{21} + \Delta t)\right]\} \\ & \times J_1\{2\theta_3 \sin\left[\omega_q (T_{21} + \Delta t)\right]\} \\ & \times e^{-t_{\rm echo}/\tau_{\rm decay}}.\end{aligned} \quad (11)$$

Here, $\theta_3$ is the pulse area of the third sw pulse and the time relative to the echo time is $\Delta t = t - 2T_{21} - T_{32}$. This signal also exhibits recoil modulation that is only a function of $T_{21}$.

Based on Ref. [34], in the presence of gravity, the phase of the three-pulse echo signal can be shown to be

$$\phi_g^{(3)} = -\frac{qg}{2}(2T_{21}^2 + 2T_{21}T_{32} + 2T_{32}\Delta t + 4T_{21}\Delta t + \Delta t^2). \quad (12)$$

Just as in the two-pulse case, this phase is proportional to the space-time area enclosed by the interferometer pathways shown in Fig. 1(b). Setting $T_{32} = 0$ reduces $\phi_g^{(3)}$ to the earlier result for $\phi_g^{(2)}$.

Once again, it is useful to decouple the expression for the signal into a part that is dependent on the timescales $T_{21}$, $T_{32}$ and a second part that is dependent on $\Delta t$ to explain the characteristics of the echo signal. Therefore we write the three-pulse echo signal as

$$E_g^{(3)} = E_{\rm D}^{(3)}(\Delta t) E_{\rm AI}^{(3)}(T_{21}, T_{32}) e^{i\phi_{\rm D}^{(3)}(\Delta t)} e^{i\phi_{\rm AI}^{(3)}(T_{21}, T_{32})}, \quad (13)$$

where

$$\phi_{\rm D}^{(3)}(\Delta t) = -qg\left[(T_{32} + 2T_{21})\Delta t + \Delta t^2/2\right], \quad (14)$$

is the Doppler phase which is dependent on $\Delta t$, and

$$\phi_{\rm AI}^{(3)}(T_{21}, T_{32}) = -qg(T_{21}^2 + T_{21}T_{32}) \quad (15)$$

is the AI phase which is dependent only on $T_{21}$, and $T_{32}$. The Doppler electric field amplitude is given by

$$E_{\rm D}^{(3)}(\Delta t) \propto E_{\rm RO} e^{-(\Delta t/\tau_{\rm coh})^2} J_1\left[2\theta_1 \sin(\omega_q \Delta t)\right], \quad (16)$$

and the AI electric field amplitude is given by

$$\begin{aligned}E_{\rm AI}^{(3)}(T_{21}, T_{32}) = & J_1\{2\theta_2 \sin\left[\omega_q (T_{21} + \Delta t)\right]\} \\ & \times J_1\{2\theta_3 \sin\left[\omega_q (T_{21} + \Delta t)\right]\} \\ & \times e^{-t_{\rm echo}/\tau_{\rm decay}}.\end{aligned} \quad (17)$$

Here, $\phi_{\rm D}$ can be varied by changing either $T_{32}$ or $T_{21}$. Once again, the Doppler phase term produces a modulation of the echo envelope due to gravitational acceleration. Since $t_{\rm echo} = 2T_{21}$ for the two-pulse AI and $t_{\rm echo} = 2T_{21} + T_{32}$ for the three-pulse AI, the functional forms of $\phi_{\rm D}^{(2)}$ and $\phi_{\rm D}^{(3)}$ are identical.

The electric field amplitudes in Eq. 2 and Eq. 11 have similar functional forms. However, the three-pulse

AI amplitude involves the additional experimental parameter $T_{32}$. This allows the freedom to record the echo envelope as a function of $T_{32}$ for an optimized value of $T_{21}$. For non-zero $T_{32}$, $\phi_{AI}^{(3)}$ is maximized if $T_{32} = 2T_{21}$.

The dashed lines in Fig. 2(b) show the Doppler electric field amplitude as predicted by Eq. 16. This shape is generated by choosing a convenient $T_{21}$ to maximize the recoil modulated signal, modeled by Eq. 17. The solid red line shows oscillations within the echo envelope due to gravity as predicted by $E_D^{(3)} e^{i\phi_D^{(3)}}$. The functional form of this expression is identical to the two-pulse result with the appropriate definition for $t_{\text{echo}}$.

The solid red line in Fig. 2(d) shows the predicted shape of the in-phase component of the signal amplitude for the three-pulse AI as a function of $T_{32}$ as predicted by $E_{AI}^{(3)} e^{i\phi_{AI}^{(3)}}$ (see Eqs. 15 and 17). This signal does not show recoil modulation and exhibits a characteristic period $\tau_v$ that is determined by $T_{21}$. The gray line in Fig. 2(d) shows the total signal amplitude $E^{(3)}(T_{21}, T_{32})$ predicted by Eq. 17 as a function of $T_{32}$. This signal exhibits a smooth decay due to signal loss associated with transit time and decoherence effects.

## III. DESCRIPTION OF EXPERIMENT

Fig. 3(a) shows a block diagram of the laser table. A Ti:Sapphire laser is used to generate light for atom trapping and interferometry. This laser is locked to a rubidium spectral line using saturated absorption and a tuning AOM. The light for atom trapping is frequency shifted 15 MHz below the $F = 2 \rightarrow F' = 3$ resonance of $^{87}$Rb or the $F = 3 \rightarrow F' = 4$ resonance of $^{85}$Rb (defined as $\omega_0$) by a dual pass acousto-optic modulator (AOM). Light from a grating stabilized external cavity diode laser (ECDL) produces repump light. This laser is resonant with the $F = 1 \rightarrow F' = 2$ transition of $^{87}$Rb or the $F = 2 \rightarrow F' = 3$ transition of $^{85}$Rb after passing through an AOM. This laser is also locked to a rubidium atomic resonance using saturated absorption. The laser can be amplitude modulated using the AOM. The trap and repump lasers are combined on a beam splitter and coupled into an angle cleaved, anti-reflection (AR) coated optical fiber. The undiffracted beam from the trapping AOM is aligned through a dual pass "gate" AOM to generate light for atom interferometry. The output of the gate AOM is also fiber coupled. The two optical fibers are routed to an optical table supporting the atom trapping and AI setups. This optical table is mounted on pneumatic vibration isolators.

The experimental setup on the optical table is shown in Fig. 3(b). The light from the trapping fiber is expanded to a beam diameter of 5 cm and directed along three orthogonal axes of a 316L stainless steel vacuum chamber. The circularly polarized trapping beams have a diameter of 3.5 cm and they are retro-reflected after passing through the chamber. Two pairs of trapping beams are oriented along 45° from the horizontal and the third trapping beam is in the horizontal plane. The chamber is maintained at $5 \times 10^{-9}$ Torr by an ion pump with a pumping speed of 270 L/s. The pump is located at a distance of 1 m from the trapping chamber to reduce ambient magnetic fields. The chamber is surrounded by three pairs of magnetic field and gradient canceling coils. A separate set of coils wound on the stainless steel chamber with tapered forms is used to generate the magnetic gradient for atom trapping. The trapping optics, vacuum chamber, anti-Helmholtz and cancelation coils, and ion pump are supported by the optical table.

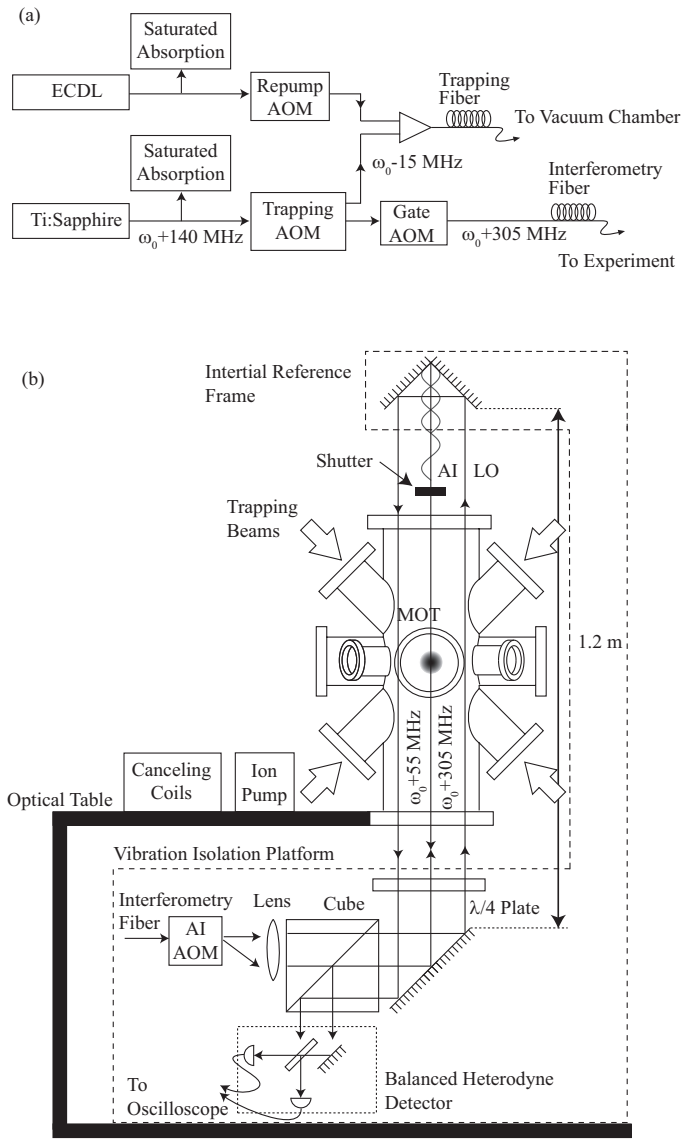

FIG. 3. Schematic of experimental setup. (a) Block diagram of laser sources and frequency control. The frequency of the laser light is defined with respect to $\omega_0$, the resonant frequency of the $^{87}$Rb $F = 2 \rightarrow F' = 3$ transition or the $^{85}$Rb $F = 3 \rightarrow F' = 4$ transition. (b) Schematic of the atom interferometry setup. The lower detection optics and the upper corner-cube reflector are anchored together and placed on a vibration isolation platform. The vacuum chamber and vibration isolation platform rest on an optical table supported by pneumatic legs. The photodiodes detect a 250 MHz beat note, which is the frequency difference between the AI and LO beams. The forms of the ion pump, canceling coils and anti-Helmholtz trapping coils are not shown.

The MOT is loaded from background vapor, with approximately $5 \times 10^8$ atoms loaded in 1 second. After the trap is fully loaded, the gradient coils are turned off in $< 0.5$ ms. The frequency of the trapping beams ($\omega_0$ - 15 MHz) is ramped by an additional 25 MHz below resonance to cool the atoms in an optical molasses. Time of flight CCD images of atoms released from the molasses [42] show that the typical temperature is 20 $\mu$K.

The light from the interferometry fiber in Fig. 3(a) is aligned through a single pass AOM operating at 250 MHz as shown in Fig. 3(b). The diffracted beam from this AOM has an average diameter of 0.8 cm and it is detuned by $\Delta \approx 55$ MHz above resonance. This excitation beam is directed along the vertical through AR coated viewports. This beam is circularly polarized and retro-reflected through the cloud by a corner-cube reflector to produce standing wave excitation. The undiffracted beam, with a frequency of $\omega_0 + 305$ MHz serves as an optical local oscillator (LO). The LO is aligned through the same optical elements as the excitation beam to minimize the impact of relative phase changes due to vibrations. The centers of the two beams are separated by 2.5 cm. The LO is physically displaced upon reflection by the corner-cube. To reduce the amount of background light entering the apparatus during the AI pulse sequence, the gate AOM is pulsed on only when the AI AOM is turned on. The excitation and LO beams are combined on a beam splitter and sent to a balanced heterodyne detector, which measures a beat signal at a frequency $\omega_{RF} = |\omega_{AI} - \omega_{LO}| = 250$ MHz. At the time of the readout pulse, a mechanical shutter with an open/close time of 1 ms [43] blocks the retro-reflection of the excitation beam to produce a traveling wave. The balanced detector consists of two oppositely biased Si photodiodes with rise-times of 1 ns and responsivities of 0.45 A/W.

The corner-cube reflector, AOM, and associated optics are anchored to a vibration isolation platform as shown in Fig. 3(b). This platform has a resonance frequency of 1 Hz and rests on the pneumatically supported optical table. The optical table is effective in suppressing vibration frequencies above 100 Hz, whereas the vibration isolation platform is much more effective in suppressing frequencies in the range of $1 - 100$ Hz. The mechanical shutter is separately anchored to the ceiling of the laboratory to reduce vibrational coupling. In this setup, critical components the experiment are isolated with the vibration isolation platform.

Digital delay generators (slaved to a 10 MHz rubidium clock with an Allan variance of $10^{-12}$ in 100 seconds) are used to produce RF pulses that drive the gate and AI AOM. High isolation RF switches ensure that the RF pulses have on/off contrast of 90 dB. The time delay of the corresponding optical pulses can be controlled with a precision of 50 ps.

The readout pulse intensity is of the order of the saturation intensity of 3.58 mW/cm$^2$ so that the entire echo envelope can be recorded without appreciably decohering the signal. The echo signal is measured by the balanced heterodyne detector in the form of a 250 MHz beat note. The beat frequency is recorded on an oscilloscope with an analog bandwidth of 3.5 GHz. The signal is mixed down to DC using the RF oscillator driving the AOM to generate the in-phase ($E_0 \cos(\phi_g)$), and in-quadrature ($E_0 \sin(\phi_g)$) components of the back-scattered electric field.

During trap loading, an attenuated excitation beam is turned on to record a 250 MHz beat note that monitors the phase changes due to motion of the optical elements. This measurement is used to re-initialize the RF phase used to mix the signal down to DC at the beginning of each repetition of the experiment. This procedure ensures that the relative phases between the excitation beam and the LO are the same at the start of the experiment. Although the LO and AI beams are strongly correlated at the beginning of the experiment, the phase uncertainty progressively increases with the timescale of the experiment. The typical repetition rate of the experiment varies between 0.8-3 Hz.

The residual magnetization of the stainless steel vacuum chamber produces a magnetic gradient that results in signal amplitude oscillations similar to those shown in Refs. [32, 34]. For this reason, the gradient cancelation coils are adjusted to obtain a smoothly decaying amplitude. For the two-pulse AI, we typically realize an overall timescale of $2T_{21} = 25$ ms. We obtain a significantly longer timescale for the three-pulse AI of typically $2T_{21} + T_{32} = 50$ ms.

## IV. RESULTS

### A. Doppler Phase Measurements

We first present data showing that the characteristics of the echo envelope can be used to extract the effective acceleration along the axis of excitation, $a$. Although the value of $a$ is dominated by $g$, we make this distinction because the experiment has significant systematic effects. Equation 5 for the two-pulse AI and Eq. 14 for the three-pulse AI show that the Doppler phase $\phi_D$ produces a similar modulation of the echo envelope for the two configurations of the AI.

If $T_{21}$ and $T_{32}$ are small, the echo envelope has a simple dispersion shape shown in Fig. 2(a), and predicted by Eq. 7. As $T_{21}$ and $T_{32}$ are incremented, the effect of gravity produces oscillations in the signal shape as shown in Fig. 4(a) and 4(b). These figures show the in-phase component of the echo envelope obtained on a single acquisition for two-pulse and three-pulse configurations of the experiment respectively. The effect of $a$ is apparent for the echo time $2T_{21} = 9.3$ ms in Fig. 4(a). In Fig. 4(b), the echo time $2T_{21} + T_{32} = 45.1$ ms with $T_{21} = 1.5$ ms. Therefore, the effect of gravity produces an increased modulation frequency within the echo envelope as predicted by $E_D^{(2)} e^{i\phi_D^{(2)}}$ for the two-pulse AI (see Eqs. 5 and 7), and $E_D^{(3)} e^{i\phi_D^{(3)}}$ for the three-pulse AI (see Eqs. 14 and 16).

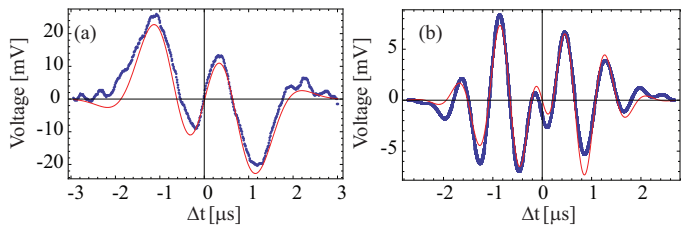

FIG. 4. (Color online) (a) Example of fit to the in-phase component of the two-pulse echo signal obtained on a single acquisition for $2T_{21} = 9.3$ ms. (b) Example of fit to the in-phase component of the three-pulse echo signal obtained on a single acquisition for $2T_{21} + T_{32} = 45.1$ ms.

For the analysis of the echo envelope, we assume that the Doppler modulation frequency across the echo envelope is a constant. The data are fit to the form

$$A(t-t_0)e^{-(t-t_0)^2/2\tau^2}\sin[\omega(t-t_0)] + D. \quad (18)$$

This fit function is based on $E_D^{(2)} e^{i\phi_D^{(2)}}$ and $E_D^{(3)} e^{i\phi_D^{(3)}}$ that define the echo envelope. All parameters except $t$ are free parameters in the non-linear least-squares fitting routine. The fit shown in Fig. 4(a) is used to infer the frequency $\omega_D^{(2)}$ given by

$$\omega_D^{(2)} = \left|\frac{\partial \phi_D^{(2)}}{\partial \Delta t}\right| = qg(2T_{21} + \Delta t), \quad (19)$$

and the fit shown in Fig. 4(b) is used to infer the frequency $\omega_D^{(3)}$ given by

$$\omega_D^{(3)} = \left|\frac{\partial \phi_D^{(3)}}{\partial \Delta t}\right| = qg(2T_{21} + T_{32} + \Delta t). \quad (20)$$

Figure 5(a) and Fig. 5(b) show the expected linear increase in frequency as a function $T_{21}$ for the two-pulse AI and as a function of $T_{32}$ for the three-pulse AI, respectively. In these graphs, the frequency is determined by the weighted average of eight repetitions, and the error bar represents the standard deviation of the distribution of these repetitions.

Based on Eq. 19, the two-pulse angular frequency is given by $\omega_D^{(2)} = qg(2T_{21}+\Delta t) \approx qg(2T_{21})$ since $T_{21} \gg \Delta t$. Similarly, based on Eq. 20, the three-pulse angular frequency is given by $\omega_D^{(3)} = qg(2T_{21} + T_{32} + \Delta t) \approx qg(2T_{21} + T_{32})$ since $T_{21}, T_{32} \gg \Delta t$. This approximation is consistent with the assumption of a constant frequency across the echo envelope. The predicted dependence of $\omega_D$ on $\Delta t$ in Eq. 19 and Eq. 20 can be measured in a longer timescale experiment that has adequate sensitivity as shown later in this section.

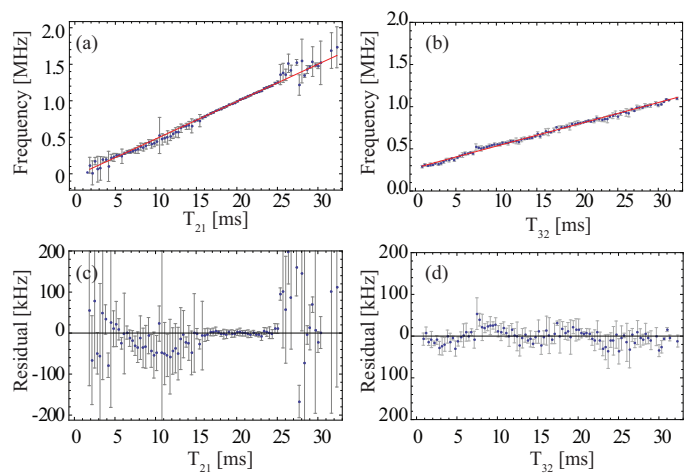

FIG. 5. (Color online) (a) Two-pulse Doppler phase measurement of $a$ by varying $T_{21}$. The slope of the line is 50.9(3) MHz/s with a corresponding statistical precision of 0.6%. (b) Three-pulse Doppler phase measurement of $a$ by varying $T_{32}$ at a fixed $T_{21} \approx 7.5$ ms. The slope of the line is 25.4(2) MHz/s, with a corresponding statistical precision of 0.8%. (c) Residuals of the two-pulse linear fit. (d) Residuals of the three-pulse linear fit. The intensity of the AI pulses is 50 mW/cm² and $\Delta = 55$ MHz. Here $\tau_1 = 0.95$ $\mu$s, $\tau_2 = 90$ ns for the two-pulse AI, and $\tau_3 = 90$ ns for the three-pulse AI. These results were obtained using $^{87}$Rb.

Linear least squares fits to the two-pulse and three-pulse data sets show that the slopes are 50.9(3) MHz/s and 25.4(2) MHz/s, respectively. The corresponding uncorrected values of $a$ are 9.93(6) and 9.91(4) m/s², respectively. These results confirm that the slope for the two-pulse pulse AI is twice the slope for the three-pulse AI. We note that the values of $a$ determined from both AI configurations exhibit a significant dependence on the fit function used to model the echo envelope. Therefore, the values of $a$ obtained from the Doppler phase are suitable for a relative comparison between the AI configurations, but not for an absolute measurement of acceleration.

The size of the statistical error is limited mainly by the relatively small temporal duration of the echo envelope, which is a few microseconds. An additional challenge is associated with fitting to the complicated signal shapes. This is illustrated by the residuals for these data sets shown in Figs. 5(c) and 5(d). Both the error bars and residuals are larger for the two-pulse AI. There are a number of factors that contribute to the characteristics of the residuals. For small $T_{21}$, the frequency of the two-pulse signal tends to zero and there are very few oscillations across the echo envelope, thereby leading to a large uncertainty. In contrast, for the three-pulse AI, $T_{21}$ was fixed at $\approx 7.5$ ms, giving rise to discernible oscillation frequencies even if $T_{32}$ is small, which leads to reduced uncertainty for this range of $T_{32}$. The signal am-plitude decreases as a function of $T_{21}$ for the two-pulse AI. Although the overall signal amplitude for the three-pulse AI is about 50% smaller, the signal decays more gradually as a function of $T_{32}$. Reduction in signal amplitude is caused by transit time losses, the Doppler shift due to falling atoms that prevents resonant two-photon excitation, and decoherence effects due to magnetic field curvature. Due to the reduced sensitivity to the last mentioned effect, the signal-to-noise ratio is generally higher for the three-pulse AI for large $T_{32}$ resulting in reduced error bars.

It should be emphasized that vibrations and magnetic field curvature affect the signal amplitude and phase for both configurations of the AI. The longest timescale for the three-pulse AI ($2T_{21} + T_{32} = 45$ ms) is attained with $T_{21} \approx 7.5$ ms, whereas the longest timescale of the two-pulse AI $T_{21} = 32$ ms. Because of the smaller $T_{21}$, we expect the three-pulse AI to be much less sensitive to these effects than the two-pulse AI, giving better fits to the echo envelope. Accordingly, the overall size of the error bars in Fig. 5(c) are larger than in Fig. 5(d).

Although the three-pulse results may appear to be more precise, the statistical precision from both AI configurations is about the same since the three-pulse frequency spans a smaller range. Another aspect of the two-pulse data is that the size of the error bars and residuals is noticeably smaller near $T_{21} = 6$ ms and 20 ms. This effect is not fully understood and we speculate that there are quiet zones in certain vibrational frequency bands.

For this discussion, we analyzed the modulation frequency of the echo envelope as a function of $T_{21}$ and $T_{32}$ to verify the predicted dependence of the Doppler phase in Eq. 5 and Eq. 14. We note that the change in modulation frequency within the echo envelope discussed in this section can also be observed by varying the onset of the AI experiment with respect to trap turn-off. In practice, this can be accomplished by varying the time of the first pulse, $T_1$, and keeping the pulse separations $T_{21}$ and $T_{32}$ fixed.

### B. Two-Pulse AI Measurement

As noted in section II, the in-phase and in-quadrature component amplitudes can be used directly to measure acceleration $a$. To obtain the in-phase component amplitude, the signal is background subtracted, and the points are squared and summed over the duration of the echo signal. This method of analysis is particularly sensitive to background subtraction. In general, fitting to the signal shape and extracting the amplitude avoids this problem if the fits are of good quality. However, as shown in Fig. 4, amplitude extraction does not produce consistently good results because of the complicated signal shape. Therefore, the squared-sum method is used to obtain the component amplitude from raw data as a function of $T_{21}$. The quadrature sum of the component amplitudes gives the total signal amplitude $E_0^{(2)}$. Each of the component amplitudes are normalized with respect to $E_0^{(2)}$ to obtain $\cos(\phi_{AI}^{(2)})$ and $\sin(\phi_{AI}^{(2)})$. This procedure is suitable for extracting $\phi_{AI}^{(2)}$ predicted by Eq. 6, but ignores the frequency variation across the echo envelope predicted by Eq. 3.

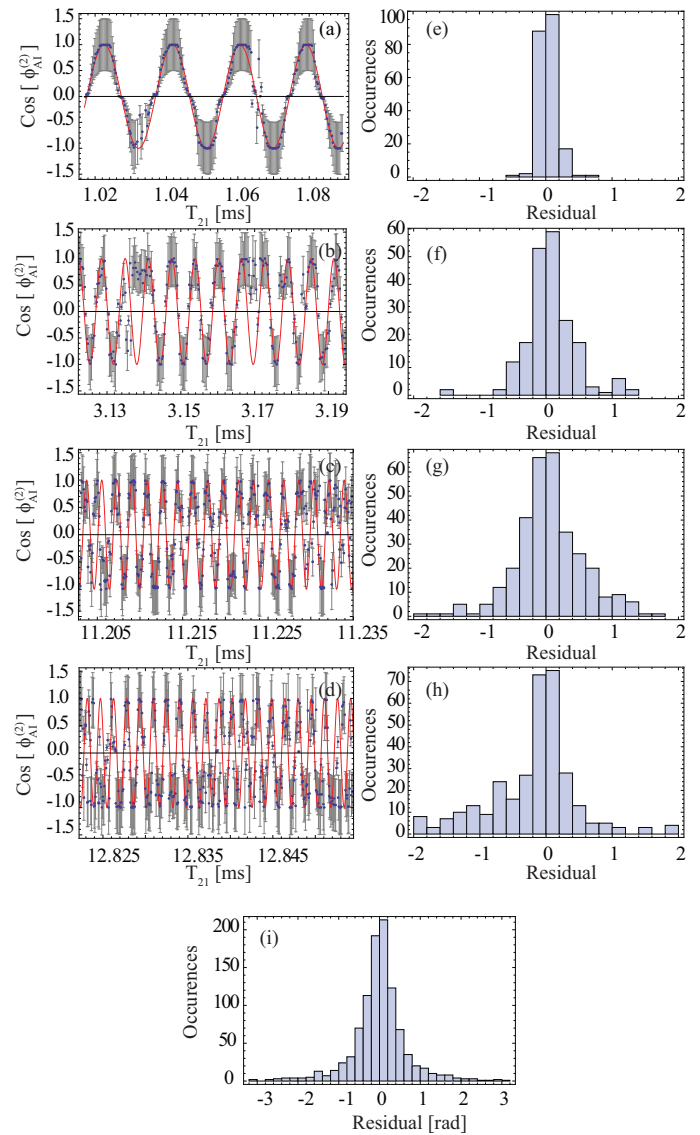

FIG. 6. (Color online) (a-d) Four observational windows showing the amplitude of the in-phase component for the two-pulse AI as a function of $T_{21}$. The angular frequency $\partial \phi_{AI}^{(2)}/\partial T_{21}$ increases linearly as a function of $T_{21}$. Here, $\tau_1 = 0.8$ $\mu$s, $\tau_2 = 90$ ns. (e-h) Histogram of the residuals. The standard deviations of the residuals for the four windows are 0.13, 0.39, 0.60, and 0.73 respectively. Here, the intensity of the AI pulses is 50 mW/cm$^2$ and $\Delta = 55$ MHz. (i) Histogram of phase residuals for the entire data set. The standard deviation is 0.7 rad. These results were obtained using $^{85}$Rb.

Figure 6 shows a measurement of $a$ using the non-linear dependence of $\phi_{AI}^{(2)}$ on $T_{21}$ as predicted by Eq. 6. Here,

the best statistical precision was obtained with the upper corner-cube reflector placed on a vibration isolation platform and the lower AI optical setup placed on separate, uncoupled vibration isolation platform, with both platforms resting on the pneumatically supported optical table. The amplitude of the in-phase component is recorded as a function of $T_{21}$ using four observational windows in a data acquisition time of one hour. Each window consists of about 200-325 points acquired in randomized order. Each data point is obtained by analyzing the echo envelope averaged over 16 repetitions. The error bars represent the standard devitation of these repetitions. The overall time window was limited to $T_{21} = 12.8$ ms because of the progressive breakdown of the periodically reset RF phase.

Figure 6 shows the expected chirped sinusoidal dependence of $\cos(\phi_{AI}^{(2)})$ on $T_{21}$. The data for the in-phase component are fit to a multi-parameter fit function of the form $\cos(qaT_{21}^2 + qv_0T_{21} + \phi_0)$ to extract $a = 9.791\ 19(8)$ m/s$^2$. This measurement has a statistical uncertainty of 8 ppm. Similarly, we obtain $a = 9.791\ 35(8)$ m/s$^2$ from the in-quadrature component which has a statistical uncertainty of 8 ppm. From a weighted average of the in-phase and in-quadrature components, we obtain $a$ with a statistical precision of 6 ppm. In this analysis, $v_0$ models a velocity parameter for the atoms, and $\phi_0$ is the initial phase of the grating with respect to the nodal point on the inertial reference frame (corner-cube reflector). The typical value of $v_0$ from the fit was $0.107(1)$ mm/s. The physical origin of $v_0$ is not clear since cloud launch does not contribute to the two-pulse AI phase. In fact, centroid tracking of the atom cloud showed launch velocities of as high as 2 mm/s along the vertical. We speculate that intensity imbalances in the two standing wave components can produce this effect.

The standard deviations of the residuals shown in Fig. 6(e-h) increase as $T_{21}$ is increased. However, the size of the residuals in all four windows is smaller than the standard deviation for a random distribution of points distributed uniformly between -1 and 1, which is 0.9 based on simulations. The residuals can be converted to phase units to enable a more effective comparison. Figure 6(i) shows a histogram of the phase residuals for the entire data set. The standard deviation of the phase residuals for the entire data set is 0.7 rad out of a total phase accumulation of $|\phi_{AI}^{(2)}| \sim 2.6 \times 10^4$ rad for $T_{21} = 12.85$ ms. The increasing size of the residuals for $T_{21} > 10$ ms illustrates the sensitivity of the two-pulse AI to vibrations and other decoherence effects such as magnetic field curvature.

Another indication that the two-pulse AI was more sensitive to vibrations is that the best statistical precision was obtained by adding an additional vibration isolation platform to support the upper reflector. In comparison, it was possible to obtain much better statistical precision for the three-pulse AI, as shown later in this section, by using a single vibration isolation platform to isolate the AI optics and corner-cube reflector in Fig. 3.

### C. Three-Pulse Measurement

We now discuss the data obtained with the three-pulse AI. The relative insensitivity of the three-pulse AI to vibrations compared to the two-pulse AI allows us to avoid the two previously mentioned analysis techniques, namely fitting to the echo envelope as well as the faster square-sum method. Instead, we use a 'slicing' technique. The instantaneous amplitude of the background subtracted data is found from a single time slice of the echo envelope as shown in Fig. 2(a). The best statistical precision was obtained with a temporal duration of 10 ns. This slice duration was chosen since there is effectively no change in the signal amplitude over this timescale. The average amplitude of each slice was determined by averaging 16 repetitions.

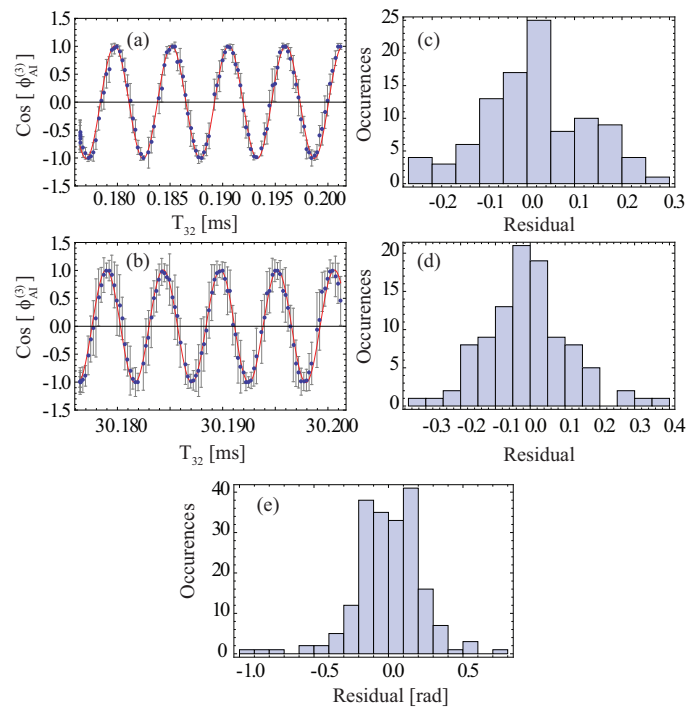

FIG. 7. (Color online) Two observational windows of the amplitude of the in-phase component for the three-pulse AI versus $T_{32}$ in the range of (a) 0.2 ms and (b) 30 ms. These data correspond to a single time slice and exhibits a constant frequency as a function of $T_{32}$. Here, $T_{21} = 7.431900$ ms. The frequency extracted from the fit is $187\ 324.75(8)$ Hz. Figures (c) and (d) show histograms of residuals for the data in Figs. (a) and (b) respectively. (e) Histogram of phase residuals for the total data set. The standard deviation is 0.2 rad. These results were obtained using $^{87}$Rb.

Figure 7(a) and Fig. 7(b) show $\cos(\phi_g^{(3)})$ as a function of $T_{32}$ for a single slice with $T_{21}$ fixed at $7.431\ 900$ ms. The in-phase and in-quadrature components $\cos(\phi_g^{(3)})$ and $\sin(\phi_g^{(3)})$ were obtained by following the same normalization protocol as the two-pulse AI. These data were recorded with 100 points in each window acquired in ran-domized order in one hour. As shown by Eq. 15, for a fixed maximum timescale $t_{echo} = 2T_{21} + T_{32}$, the greatest phase accumulation that can be obtained for the three-pulse AI by varying $T_{32}$ is realized if $2T_{21} = T_{32}$. Since the largest $t_{echo} = 45$ ms, the ideal value of $T_{21} \approx 11$ ms. However, we choose to operate with $T_{21} \approx 7.5$ ms to reduce the effect of vibrations and decoherence mechanisms. As predicted by Eq. 15, the signal exhibits a single frequency that can be precisely determined using two widely-spaced observational windows.

If the frequency change across the echo envelope predicted by Eq. 12 is ignored, the frequency for a single slice can be written as

$$\omega_{AI}^{(3)} = \left| \frac{\partial \phi_{AI}^{(3)}}{\partial T_{32}} \right| = q a T_{21}, \quad (21)$$

which allows $a$ to be determined. Using $T_{21} = 7.431900$ ms, and $q = 16\ 105\ 651.65\ [44]$ rad/m, we obtain $a = 9.833\ 245(4)$ m/s$^2$ from the in-phase component, which represents a statistical precision of 0.4 ppm. This statistical precision can be compared to the 6 ppm statistical uncertainty for the two-pulse AI.

The enhancement in precision can be attributed to several effects. Firstly, the analysis involves fitting to a single frequency in the absence of recoil modulation. Secondly, the measurement time scale has also been significantly extended in comparison to the two-pulse AI since these data represent a total timescale $2T_{21} + T_{32} = 45$ ms while limiting the value of $T_{21}$ to $\approx 7.5$ ms. Therefore there is reduced sensitivity to the effects of magnetic curvature and vibrations. The insensitivity to these effects also leads to a more gradual decay of the signal amplitude. As a result, we obtain a similar standard deviation of the residuals ($\approx 0.11$) in each observational window. The standard deviation of phase residuals for the entire data set is 0.2 rad as shown in Fig. 7(e). In comparison, the overall standard deviation is 0.7 rad for the two-pulse AI. Indeed, the slicing technique cannot be expected to work without good phase stability.

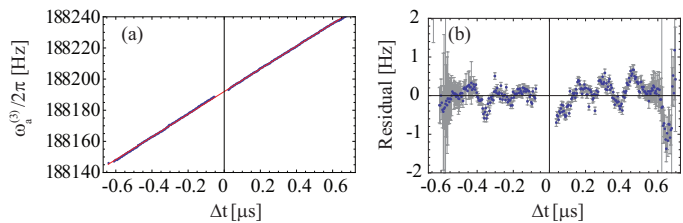

FIG. 8. (Color online) (a) Measurement of $\omega_a^{(3)}$ across the echo envelope using 200 time slices. Here we use $T_{21} = 7.431\ 900$ ms. The intercept of the line is $188\ 192.095(17)$ Hz giving an uncorrected value of $a = 9.877\ 445\ 8(9)$ m/s$^2$. The corresponding statistical error is 90 ppb. Similar analysis of the in-quadrature component reduces the combined statistical error to 75 ppb. (b) Residuals of the linear fit. These results were obtained using $^{87}$Rb.

The slicing technique allows the frequency across the echo envelope predicted Eq. 12 to be observed. By determining the frequencies of all time slices across the echo envelope, a further improvement in statistical uncertainty is achieved. Figure 8(a) shows the frequency of each time slice as a function of $\Delta t$. For this analysis, the echo envelope is divided into 200 slices, each with a duration of 10 ns. The data confirm the linear dependence of the angular frequency on $\Delta t$ predicted by

$$\omega_a^{(3)} = \left| \frac{\partial \phi_a^{(3)}}{\partial T_{32}} \right| = qa(T_{21} + \Delta t), \quad (22)$$

with $\phi_a^{(3)}$ given by Eq. 12 by replacing $g$ with $a$. Each data point in Fig. 8(a) has a typical error bar of 1 ppm (one of the best data sets is presented in Fig. 7). The reduction in error in comparison to the two-pulse AI is a result of the 30 ms time scale spanned by the two observational windows. With this improved sensitivity, it is possible to observe the change in frequency with $\Delta t$ across the echo envelope. In contrast, for the data in Fig. 4, we assume a constant frequency across the echo envelope since the observational window is only a few microseconds long and the measurement does not have the desired sensitivity.

For the data in Fig. 8(a), the limited timescale of the echo envelope and the scatter lead to an overall statistical error in the slope that is appreciable (600 ppm) despite the relatively small statistical error in each of the points (1 ppm). The scatter is attributed to magnetic effects described in the next section. However, the error in the frequency intercept is much more tightly constrained since the data are closely clustered near $\Delta t = 0$. Based on Eq. 22, the frequency intercept is $qaT_{21}$. From the linear fit in Fig. 8(a), we determine this frequency intercept at $\Delta t = 0$ to be $188\ 192.095(17)$ Hz. Using the values of $q$ and $T_{21}$, we obtain $a = 9.877\ 445\ 8(9)$ m/s$^2$, which represents a statistical precision of 90 ppb. A weighted average of the measurements from the in-phase and in-quadrature components gives a combined statistical precision of 75 ppb. Figure 8(b) shows the residuals to the straight line fit in Fig. 8(a). The residuals increase in size in the regions where the signal is small such as in the extremities and in the vicinity of $\Delta t = 0$.

We have also investigated the scaling of $\omega_a^{(3)}$ on $T_{21}$ predicted by Eq. 22. These results are shown in Fig. 9. For these data, the maximum timescale $2T_{21} + T_{32}$ is fixed at 50 ms, and the time window over which the atoms are sensitive to accelerations, $T_{21}$, was varied. For each value of $T_{21}$, the frequency is determined from the intercept, as in Fig. 8(a). The data are fit to a linear function, and the slope was found to be $qa/(2\pi) = 25.41(6)$ MHz/s. The corresponding acceleration is $a = 9.91(2)$ m/s$^2$. Although each data point has a statistical uncertainty corresponding to 90 - 150 ppb, the uncertainty in the slope is $\sim 0.2\%$ because the data do not follow a perfectly linear trend—as indicated by the residuals shown in Fig. 9(b). The nature of this trend is evidence of a number of systematic effects that we discuss in section V, such as the curvature of the residual magnetic field and the variation in the initial velocity of the atoms for different repetitions of the experiment. These sources of error cannot be controlled in the current apparatus.

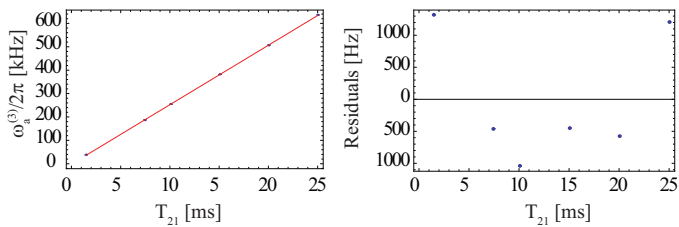

FIG. 9. (Color online) (a) Change of $\omega_a^{(3)}$ as a function of $T_{21}$. The vertical error bars (90-150 ppb) are too small to be seen. The slope of the linear fit is $qa/(2\pi) = 25.41(6)$ MHz/s. The corresponding value of $a$ is 9.91(2) m/s$^2$. (b) Residuals of the linear fit to the data shown in (a). These results were obtained using $^{87}$Rb.

## V. DISCUSSION OF RESULTS AND SYSTEMATIC EFFECTS

### A. Summary

Table I summarizes the values of $a$ obtained using the different techniques outlined in section IV. The value of $a$ extracted from the data has been corrected for two controllable systematic effects. The first is the angle of the excitation beam $\vartheta_\perp$ with respect to the vertical, which produces a correction $a' = a/\cos\vartheta_\perp$. The typical value of $\vartheta_\perp$ for the results in Table I is 15 mrad. Since this correction modifies $a$ at the level of 110 ppm, it only impacts entries three and four. The second correction only applies to the three-pulse velocity dependent measurements in Fig. 7 and Fig. 8, which is the fourth entry in Table I. For this measurement, the excitation pulse sequence was applied 88 $\mu$s after the extinction of the trapping beams. Thus, the velocity acquired during this delay time has to be considered when extracting $a$. This correction is explained in greater detail in the latter part of this section. The angle correction is also too small to affect the value of $a$ for the last entry. Here, the onset time correction impacts all points in the same manner, and therefore does not change the slope.

The measurements in Table I can be compared to the absolute value of $g$ determined by a falling corner-cube optical interferometer (Scintrex model FGL) located in the laboratory. The baseline value of $g = 9.804\ 165\ 15(5)$ m/s$^2$ was obtained with an overall precision of 5 ppb (sum of statistical and systematic uncertainties) by subtracting the effect of tides and the Earth's gravity gradient which was assumed to be 300 ppb of $g$ per meter. As previously noted in this section IV, the value of $a$ based on the determination of the Doppler phase in the first two entries is sensitive to the nature of the fit function. Therefore, great caution should be exercised in using these values for absolute measurements of $g$. The high precision measurements, in lines three and four, agree with the baseline value of $g$ at the level of 0.1%, whereas the last entry, which is a three-pulse AI measurement, agrees to within 1.1%. The errors for the three-pulse AI represent a satisfactory level of agreement since the estimate of systematic effects is 2.2% as discussed later in this section. For the two-pulse AI, the estimate of the systematic effects is 630 ppm, which is insufficient to explain the error. However, as explained in this section, the estimate is error prone as the major systematic effects cannot be independently controlled.

| Method | Measured $a$ [m/s$^2$] | Corrected $a$ [m/s$^2$] | Error |
|---|---|---|---|
| Two-Pulse Echo Envelope | 9.93(6) | 9.93(6) | 1.3% |
| Three-Pulse Echo Envelope | 9.91(4) | 9.91(4) | 1.1% |
| Two-Pulse Amplitude | 9.791 19(8) | 9.792 29(8) | 0.1% |
| Three-Pulse Amplitude | 9.877 445 8(9) | 9.816 237 2(9) | 0.1% |
| Slope from Fig. 9 | 9.91(2) | - | 1.1% |

TABLE I. Summary of measured values of $a$ from the in-phase component of the signal using the methods described in section IV. The error is calculated relative to a baseline value of $g$ in the laboratory measured with a falling corner-cube gravimeter. The corrections pertain to controllable systematic effects due to the verticality which affects all results and initial velocity acquired during $T_1$, which affects only the three-pulse results.

| Systematic | Estimate | Shift |
|---|---|---|
| Verticality | $\vartheta_\perp = 15$ mrad | +110 ppm |
| Index of Refraction | $n = 0.9999955$ | -4.5 ppm |
| Diffraction | $\vartheta_{\text{Div}} = 0.8$ mrad | $\pm 0.35$ ppm |
| Magnetic gradient | $\nabla B = 8$ mG/cm | $\pm 520$ ppm |
| Magnetic curvature | $\nabla^2 B = 4.3$ mG/cm$^2$ | $\pm 340$ ppm |
| Three-Pulse Time Offset | $T_1 = 88$ $\mu$s | -2000 ppm |
| Three-Pulse $T_{21} = 7.432$ ms Launch Velocity | $v_L = 2$ mm/s | $\pm 22\ 000$ ppm |

TABLE II. Summary of dominant systematic effects on $a$ that limit the accuracy of the experiment.

### B. Systematic Effects

Table II shows the typical values of the main systematic effects that must be used to correct $a$. Based on Eq. 3 and Eq. 12, the error in $a$ is linearly dependent on the error in $q$. The corner-cube reflector ensures that the angle between the two traveling wave components of each sw pulse deviates from 180° by no more than 3 arcseconds, corresponding to a 0.1 ppb change in $q$. This is negligible compared to other sources of error. Since the standing waves are aligned to within 15 mrad of the vertical, there is a larger correction of 110 ppm associated with $a$. The index correction, which impacts $q$, is dependent on both the sample density and the detuning of the excitation [45]. We estimate this correction to be -4.5 ppm for the typical sample density of $4.4 \times 10^9$ atoms/cm$^3$ and a detuning of 55 MHz. The excitation beam collimation was restricted by the size of the optical breadboard on the vibration isolation platform in Fig. 3 so the divergence of the laser beam resulted in a correction of 350 ppb.

The magnetic gradient in the vicinity of the atom cloud is significantly influenced by the magnetized stainless steel vacuum chamber. Gradient canceling coils are used to reduce the average gradient so that the timescale of the experiment is maximized. Based on references [32] and [34], it is known that the echo signal exhibits amplitude oscillations with a magnetic phase

$$\phi_M = \frac{q m_F g_F \mu_b \nabla B}{2M} \Upsilon \quad (23)$$

in the presence of a magnetic gradient $\nabla B$, where $\Upsilon = T_{21}^2$ for the two-pulse AI, and $\Upsilon = T_{21}^2 + T_{32} T_{21}$ for the three-pulse AI, $g_f$ is the Landé g factor, and $\mu_b$ is the Bohr magneton. Since it was not possible to directly measure the magnetic gradient at the location of the atoms, we used the measured frequency of amplitude oscillations to estimate the gradient traversed by atoms during the AI pulse sequence. This estimate of 8 mG/cm suggested that the correction to $a$ should be of order 520 ppm for an atom in the $m_F = 2$ sublevel.

The gradient systematic can be predicted on the basis of numerical simulations for arbitrary $m_F$ population distributions. Based on reference [41, 46], it is possible to describe the total backscattered electric field as a superposition of electric field amplitudes from individual sublevels. These contributions are weighted by pulse areas and Clebsch-Gordan coefficients. For the three-pulse AI involving $^{87}$Rb, the simulations confirm that a gradient of 8 mG/cm will result in a maximum correction of 520 ppm for atoms pumped into the $m_F = 2$ sublevel and that other population distributions produce smaller effects.

Another systematic effect arises from the curvature of the magnetic field in the vicinity of the atoms. The origin of the field curvature is difficult to isolate since it has contributions from the canceling coils, magnetized vacuum chamber, and ion pump magnets. We now estimate this effect based solely on the properties of the canceling coils. From the geometry of the canceling coils, the curvature of the magnetic field across the atoms is estimated to be 4.3 mG/cm$^2$. The correction can be estimated based on Eq. 23, so that

$$\Delta g = \frac{m_F g_F \mu_b (\nabla^2 B \Delta z)/M}{g}. \quad (24)$$

Using the drop time of 50 ms, we obtain the drop height of $\Delta z = \frac{g}{2}(0.050)^2 = 1.2$ cm, which gives a correction of 340 ppm.

In the presence of magnetic field curvature, the gradient sampled by the interfering momentum states will have a spatial variation. The phases of the momentum states change as a function of spatial location resulting in systematic correction to $a$. The data shown in Fig. 9 exhibits a statistical error in the slope of $\pm 0.2\%$ (which represents a variation in $a$). Since the total timescale $2T_{21} + T_{32}$ was fixed at 50 ms, the momentum states sampled different spatial regions as $T_{21}$ was varied. Therefore it is possible that the observed statistical error is influenced by the curvature and gradient corrections.

### C. Velocity Effects

We now discuss velocity dependent effects on the two AI configurations. Although an initial velocity will affect the Doppler phases of both AIs, the measurement of $\phi_{AI}^{(2)}$ based on recording the signal amplitude in Fig. 6 is independent of the initial velocity of the sample. In contrast, the three-pulse measurements in Fig. 7 and Fig. 8 are sensitive to the initial velocity of the sample at the beginning of the pulse sequence.

For the three-pulse AI, there was an overall onset time of $T_1 = 88$ $\mu$s before the start of the AI pulse sequence with respect to the end of the molasses cooling stage. Since the atoms have acquired an initial velocity during this delay time because of gravity, there is a systematic offset in the frequency measured for every slice. To understand this effect, we note that the AI frequency for a single point on the echo envelope is given by Eq. 21. To include the effect of the envelope, it is necessary to use the Doppler frequency in Eq. 20. This equation is in turn modified by the effect of the onset time $T_1$ so that

$$\omega_D^{(3)} = -qg(T_1 + T_{21} + T_{32} + T_{21} + \Delta t). \quad (25)$$

This effect leads to a -2000 ppm shift in the value of $a$ in Table II for the three-pulse AI.

The launch velocity of the cloud, $v_L$, is another parameter that affects the three-pulse AI results. In this case, the frequency $\omega_D^{(3)}$ in Eq. 25 is modified by a constant shift of $q v_L$. Typically $v_L$ is sensitive to a number of effects such as power imbalances in the laser beams, imperfect circular polarization of the trapping beams, and background magnetic fields. The CCD camera method of measuring the trap temperature monitors both the spatial profile and centroid of the atom cloud [42]. Although the spatial position of the cloud cannot be measured in real-time during the AI experiment, periodic checks showed that $v_L$ along the vertical direction can vary by as much as $\pm 2$ mm/s over several hours. For the three-pulse AI, with $T_{21} = 7.432$ ms, this effect produces a correction of $\pm 2.2\%$. Since the data in Fig. 9 was taken over 10 hours, we expect this effect to impact both the statistical error in the slope of $\pm 0.2$% as well as the absolute error in the value of $a$ extracted from the slope of 1.1%. The importance of measuring and controlling $v_L$ is likely to be one of the crucial considerations for accurate, high-precision measurements of $g$ using this technique.

| Systematic | Estimate | Shift |
|---|---|---|
| Counterpropagation | $\theta = 3$ arcseconds | -0.1 ppb |
| AC Stark Shift | I = 100 mW/cm$^2$ | -25 ppb |
| Laser Linewidth | 1 MHz | $\pm 3$ ppb |
| Zeeman Shift | 100 mG | $\pm 0.4$ ppb |
| Wavefront Curvature | $w_0 = 3$ mm | $\pm 3$ ppb |

TABLE III. Summary of estimates of systematic effects that are below the current level of statistical precision. Here we assume a typical background magnetic field, a collimated beam with intensities and detunings used in the experiment.

In summary, the quadrature sum of the systematic errors is 2.2% for the three-pulse AI and 630 ppm for the two-pulse AI. These corrections satisfactorily explain the discrepancy between the baseline value of $g$ and the value of $a$ shown in Table I for the three-pulse AI. For the two-pulse AI, the estimated systematic corrections are not sufficient to explain the discrepancy. We attribute this discrepancy to the lack of independent control over the magnetic gradient and curvature effects, and the inability to directly measure the variation of the magnetic field in the current setup. Other effects that have not been characterized include the dependencies on pulse widths. Furthermore, the effect of the read-out pulse intensity, which can affect the slope of the straight line in Fig. 8, has also not been characterized. Although the major sources of systematic errors have been described, a precision measurement of $g$ using echo interferometers will require the control of several other systematic shifts. Some of these effects, estimated under well controlled conditions, are summarized in Table III. The relatively small sizes of these effects suggest that the main challenge will be associated with suppressing the effects in Table II.

## VI. FUTURE WORK

We now discuss improvements to the experiment that can be realized in a new setup with the goal of achieving a statistical precision of 1 ppb with adequate control over systematic effects. Such an effort is currently underway.

The Raman AI in Ref. [15] and the hybrid Raman AI based on large momentum transfer Bragg pulses in reference [29] relied on state selection to a $m_F = 0$ sublevel, measurement timescales of 100 ms, and passive vibration isolation to achieve statistical precision of a few ppb in 1000 s of averaging. In comparison, the statistical precision achieved in this work is limited both by the timescale due to the magnetized apparatus, low signal-to-noise ratio due to the photodiode based detection technique, and the overall phase stability.

In measurements of magnetic gradients [34] and atomic recoil [38], we have successfully used echo AIs and a non-magnetic apparatus to achieve measurement timescales of $\approx 130$ ms with the two-pulse AI, and $\approx 220$ ms for the three-pulse AI without magnetic state selection. These timescales were realized due to better control of magnetic effects and the use of chirped standing wave pulses to compensate for Doppler shifts. The non-magnetic apparatus used in these experiments reduced intensity imbalances between trapping beams resulting in improved molasses cooling and a ten-fold increase in atom number compared to this work. The well controlled magnetic environment in the glass cell has also allowed atoms to be routinely cooled to to less than 5 $\mu$K thereby limiting cloud expansion. Such an apparatus will also allow much better control of the launch velocity of the atom cloud. Additionally, a gated PMT led to a five-fold increase in signal-to-noise ratio showing that an overall 50-fold increase in signal size is readily achievable. However, these experiments did not require vibration isolation since they measured only the signal intensity, and are therefore unsuitable for measurements of $g$. The glass cell results suggest that a vibration stabilized, non-magnetic apparatus, and a heterodyne detection system using PMTs can result in the suitably long timescales and adequate signal-to-noise ratio required to achieve competitive measurements of $g$ using echo AIs.

If atoms are state selected to a $m_F = 0$ level, a three-fold loss in signal-to-noise-ratio is expected. Therefore the overall improvement in signal-to-noise is expected to be a factor of 17 with respect to the current work. Magnetic state selection into an $m_F = 0$ sublevel is also expected to increase the timescales of both AI configurations from the current timescale in the glass cell (130 ms for the two-pulse AI, 220 ms for the three-pulse AI) to the transit time limit of $\approx 300$ ms with a 2.54 cm beam diameter. This timescale is similar to those achieved using Raman AIs [3, 20, 47]. It may be possible to compensate the expected three-fold loss of signal strength due to state selection by pre-loading the atoms into a one dimensional optical lattice as in recent echo experiments [48].

Due to operational constraints, the results in this paper were obtained with passive vibration isolation of only critical components of the apparatus. To reduce phase error in such a setup, it is desirable to passively isolate the entire apparatus using a vibration isolation platform that supports all elements of the experiment including the ion pump, vacuum chamber and optics. Under these conditions, the phase error is expected to be reduced from the current level of 0.7 rad to 0.01 rad for the two-pulse AI based on the results in reference [29]. An improved apparatus can also use a Michelson interferometer to measure the relative motion between the corner-cube reflector and optics platform. Additionally, the motion of the corner-cube reflector that serves as the inertial reference frame will be actively stabilized using an accelerometer and solenoid actuator. Under these conditions, the phase error is expected to be further reduced.

Simulations of the two-pulse and three-pulse AIs with the current signal-to-noise ratio and phase noise closely match the results in this paper. These results can be extended based on the scaling laws for $\phi_{AI}^{(2)}$ and $\phi_{AI}^{(3)}$, the expected 17-fold increase in signal-to-noise ratio for a glass cell experiment that utilizes atoms selected in the $m_F = 0$ sublevel, the expected reduction of phase noise to 0.01 rad, and a total experimental timescale of 300 ms. Under these conditions, the simulated statistical precision is 0.3 ppb for the two-pulse AI and 0.6 ppb for the three-pulse AI. A further statistical enhancement, as in Fig. 8 should also be possible by exploiting the data in the echo envelope. With active stabilization of the inertial reference frame, it is likely that further improvements in the statistical precision are also achievable.

Another consideration for a long timescale echo interferometer is the necessity of maintaining the two-photon resonance condition to compensate for the gravity induced Doppler shift. In the long-timescale echo experiments in references [34, 38], both traveling wave components were chirped, whereas in reference [29], only one traveling component was chirped at twice the required rate. In this work, the standing wave pulses were not chirped since the pulse duration of $50 - 100$ ns provided sufficient bandwidth to maintain the two-photon resonance condition. A model for the two-photon transition probability as a function of the pulse width confirms these expectations. For our pulse parameters, this model shows that although the resonance condition is maintained for timescales of $> 100$ ms, the signal size is appreciably reduced. An increase in signal size can be achieved by using longer pulses and compensating for the gravitational Doppler shift by chirping the frequency.

The index of refraction systematic can be addressed by working at much larger detunings. However, such an experiment will require more power in the excitation beams to achieve higher atom-field coupling strengths and maintain the same pulse area. The development of a new family of high power fiber amplified lasers [49, 50] makes it practical to achieve the desired intensities. Since fiber lasers can have a linewidth as low as 50 kHz, the linewidth systematic in Table III can be reduced to below the desired level. Another possibility is to operate the AI at a detuning where the combined index correction due to all transitions in the $F = 2 \to F' = 1, 2, 3$ manifold of $^{87}$Rb or the $F = 3 \to F' = 2, 3, 4$ manifold of $^{85}$Rb sum to zero [38]. However, studies remain to be carried out to determine the extent of suppression of the index correction.

Decreasing the wavefront curvature systematic can be accomplished by increasing the size of the beams in the optical setup. Such a setup will allow the use of more collimated beams and reduce diffraction. This change will require a larger experimental optical breadboard and a corresponding increase in the load capacity of vibration isolators, which can be accomplished with adequate resources. It will be possible to reduce the systematic effect due to the verticality of the laser beams to below 1 ppb by adopting a procedure used in industrial gravimeters. Here, the vertical beam is aligned by relying on the horizontality of the center of a meniscus of a high-index liquid.

The three-pulse echo experiment will require better control of the launch velocity and a more precise characterization of any onset time delays. Recent experiments in a glass cell [34, 38, 51] indicate that the molasses cooling is highly efficient, routinely producing temperatures of $\approx 1$ $\mu$K. These results suggest that imbalances in the intensities of trapping beams and magnetic field effects are well controlled. Under these conditions, we expect that the effect of $v_L$ can be characterized by launching atoms by ramping detuning of trapping beams.

Another priority will be the reduction in measurement time, which can be achieved with a higher signal-to-noise ratio by recording under-sampled fringe patterns [52]. Optimizing data acquisition with respect to the number of fringes recorded and the distribution of points within an observation window will need to be investigated.

If all the statistical errors are well controlled, is it possible to "dial up" the appropriate pulse parameters and measure the displacement of the maximum of the fringe pattern to monitor relative changes in $g$. The next generation experiment in a glass chamber will incorporate all the aforementioned improvements. Such an experiment will allow tidal variations to be measured in real-time. A particularly valuable method for analyzing systematic effects will be to compare data from the cold atom experiment with real-time data from a portable industrial gravimeter as in Ref. [13].

## VII. CONCLUSION

In summary, we have developed and explored techniques for measuring $a$ using two-pulse and three-pulse echo interferometers. Analysis of the Doppler phase oscillations of the echo envelope resulted in measurements statistically precise to 0.5%. Experiments with the amplitude and phase of the two-pulse AI yielded a statistical uncertainty of 6 ppm. Experiments with the three-pulse AI with a drop height of 1.2 cm have demonstrated the best statistical precision of 75 ppb. This measurement surpasses the statistical precision attained using the Bloch oscillation technique [27], but is currently not competitive with the precision of Raman AIs [3, 20, 21, 29].

Although both the two-pulse and three-pulse echo experiments have recorded significant improvements in statistical precision compared to the 100 ppm measurement in Ref. [30] and 15 ppm measurement completed over several hours in Ref. [32], the timescale of the experiments described here was principally limited by the magnetized vacuum chamber. The magnetization of the chamber was also responsible for the dominant sources of systematic error. Based on the improved signal-to-noise ratio obtained in recent echo experiments that utilized a non-magnetic apparatus [34, 38], we have projected a measurement sensitivity of 0.6 ppb for the three-pulse AI and 0.3 ppb for the two-pulse AI. Such an experiment can be realized in a non-magnetic apparatus with active stabilization of the inertial reference frame. However, more work will be required to verify that the systematic errors can be reduced to the level of the statistical error.

## ACKNOWLEDGMENTS

This work was supported by the Canada Foundation for Innovation, Ontario Innovation Trust, Natural Sciences and Engineering Research Council of Canada, Ontario Centres of Excellence, the United States Army Research Office and York University. We thank Itay Yavin of McMaster University for helpful discussions. We thank Tycho Sleator of New York University for generously lending crucial components of a Ti:Sapphire laser. We thank Wendy Taylor of York University for the periodic loan of a high bandwidth oscilloscope.